# Lattice Gas Prediction is P-complete

Cristopher Moore[1] and Mats G. Nordahl[2]

[1] Santa Fe Institute, 1399 Hyde Park Road, Santa Fe, New Mexico 87501
e-mail: **moore@santafe.edu**
[2] Institute of Theoretical Physics, Chalmers University of Technology, S-412 96
Göteborg, Sweden
e-mail: **tfemn@fy.chalmers.se**

**Abstract.** We show that predicting the HPP or FHP III lattice gas for finite time is equivalent to calculating the output of an arbitrary Boolean circuit, and is therefore **P**-*complete*: that is, it is just as hard as any other problem solvable by a serial computer in polynomial time.

It is widely believed in computer science that there are *inherently sequential* problems, for which parallel processing gives no significant speedup. Unless this is false, it is impossible even with highly parallel processing to predict lattice gases much faster than by explicit simulation. More precisely, we cannot predict $t$ time-steps of a lattice gas in parallel computation time $\mathcal{O}(\log^k t)$ for any $k$, or $\mathcal{O}(t^\alpha)$ for $\alpha < 1/2$, unless the class **P** is equal to the class **NC** or **SP** respectively.

## 1 Introduction

Given the initial conditions of a $d$-dimensional cellular automaton, suppose we want to know the state at a site $t$ time-steps in the future. We can do this in $\mathcal{O}(t^{d+1})$ steps on a serial computer, or $\mathcal{O}(t)$ steps on a parallel one, simply by simulating the CA explicitly and filling in the light-cone above the site in question. Thus CA PREDICTION is in the class **P** of problems solvable by a serial computer (such as a deterministic Turing machine) in polynomial time.

For large $t$, it would be very nice if a short-cut were available; if instead of having to simulate every step of the CA, we could skip over a lot of it and get the answer in *polylogarithmic* time, i.e. $\mathcal{O}(\log^k t)$ steps for some $k$, on a parallel computer with a polynomial number of processors. This would put CA PREDICTION in the class $\mathbf{NC} \subset \mathbf{P}$ of problems solvable by circuits of polylogarithmic depth and polynomial size.

A problem is *complete* for a complexity class if all other problems in that class can be reduced to it; thus if it could be solved quickly, so could all other problems in its class. Just as **NP**-complete problems are believed to require a super-polynomial amount of search, **P**-complete problems are believed to be *inherently sequential*, so that the work needs to be done in step-by-step order and cannot be efficiently parallelized. Unless $\mathbf{NC} = \mathbf{P}$ (which would be almost as surprising to computer scientists as if $\mathbf{P} = \mathbf{NP}$), then, **P**-complete problems cannot be solved in polylogarithmic parallel time.

Some non-linear CA's with certain algebraic properties can be predicted in **NC** [9, 10], as can the Lorentz lattice gas of one particle bouncing off of fixed scatterers [13]. But CA PREDICTION is **P**-complete in general since CA's exist that can perform universal computation [5, 7]. Other cellular automata and lattice systems that have been shown to be **P**-complete include majority-voting CA's, single spin-flip Ising dynamics, sandpiles, fluid invasion, and diffusion-limited aggregation [8, 11, 12].

These systems are **P**-complete because the following problem, called CIRCUIT VALUE, can be reduced to them: given a Boolean circuit, i.e. a directed graph whose nodes are AND, OR and NOT gates, and given the truth values of its inputs, is its output true or false? This is **P**-complete since any deterministic Turing machine computation of length $t$ can be converted to a Boolean circuit of depth $\mathcal{O}(t)$; it seems inherently sequential, since it is hard to imagine how one could calculate the output of an arbitrary circuit without going through it level-by-level.

Then the proofs in [8, 11, 12] work by showing that we can build "wires" to carry truth values, and AND and OR gates to connect them, so that the future state of a particular site corresponds to the output of the circuit.

In this paper, we show that the same can be done for lattice gases in both the HPP and FHP III rules, so that LATTICE GAS PREDICTION is **P**-complete. Thus, unless **NC** = **P**, lattice gases in two or more dimensions cannot be predicted in polylogarithmic parallel time.

We can draw the finer distinction of whether a polynomial speedup is possible in parallel, i.e. whether $t$ time-steps can be predicted in $\mathcal{O}(t^\alpha)$ parallel time for some $\alpha < 1$. The class of problems for which this is possible is called **SP** for *semi-efficient* parallel computation [17]. Using the theory of *strict* **P**-*completeness* [1], we show that $\alpha \geq 1/2$ for lattice gases unless **SP** = **P**. We conjecture that this lower bound can be improved to $\alpha = 1$, in which case it is impossible to predict lattice gases faster than by explicit simulation.

Finally, we end the paper with a discussion of the relevance, or lack of it, of the computational complexity of lattice gases to that of continuum hydrodynamics and the Navier-Stokes equations.

We recommend [5, 15] for introductions to **NC**, **P**, and **P**-completeness.

## 2   The HPP lattice gas

The simplest lattice gas is the HPP rule on the square lattice [6]. Unfortunately, its continuum limit is non-isotropic; it has an infinite number of unphysical conserved quantities, namely the total horizontal (vertical) component of momentum in each row (column). Thus it is not a good rule for simulating hydrodynamics. However, it has the basic characteristics of a lattice gas, so we will address it first.

The rule is shown in figure 1. There is only one kind of collision where interaction takes place, namely when two particles collide head-on; the outgoing particles are then rotated by 90°.

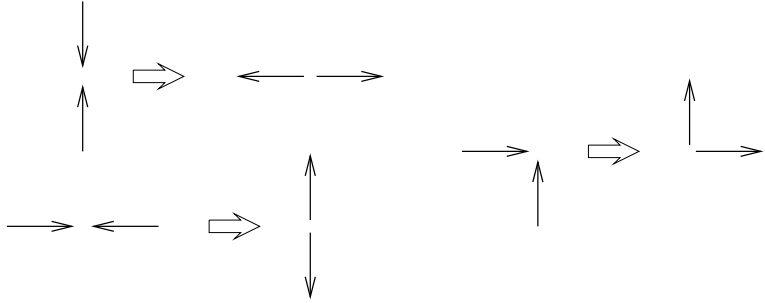

**Fig. 1.** Two-particle collisions in the HPP lattice gas. Particles colliding head-on are rotated 90°, while particles colliding at 90° don't interact.

Our proof is essentially graphical. Our truth values will be carried by upward-moving particles or "wires," which are present if true and absent if false. Then we rely on the following facts:

1.) Wires can be shifted or delayed.

2.) Wires can be split into two carrying the same truth value.

3.) Two synchronized wires (with the same $y$-coordinate) can be combined by an AND gate.

4.) Wires can be negated.

5.) The gates and wires can be arranged in such a way that the "exhaust" particles created by collisions do not interact with the wires.

(1), (2) and (3) are shown in figure 2. The AND gate, for instance, works as follows: particles $a$ and $b$, if they are present, collide with downward-moving particles (shown dashed) to produce sideways-moving ones. If both $a$ and $b$ are present, these sideways-moving particles collide to produce vertical particles; we take the upward-moving one to be $a \wedge b$.

(4) is shown in figure 3. Here, two vertical particles (dashed) collide if $a$ is not present, and their output collides with a horizontal particle, producing a vertical particle $\overline{a}$. If $a$ is present, it collides with the upper dashed particle first, preventing this from happening, and $\overline{a}$ is not produced.

Finally, we can show (5) as follows. Horizontal exhaust particles don't interact with vertical wires because 90° collisions are non-interacting, and as long as no two gates have the same $y$-coordinate, no two horizontal particles will collide to create vertical exhaust. To prevent vertical exhaust particles from colliding with the wires or with each other, we make sure that the gates have different $x$-coordinates from each other and from the wires.

Then we can implement any Boolean circuit, first by converting its gates to AND gates and negations (using De Morgan's law, $a \vee b = \overline{\overline{a} \wedge \overline{b}}$) and then taking one AND gate at a time, delaying one wire so that the two are synchronized, and combining them as in figure 2. If the number of gates is $n$, clearly the number of particles required is linear in $n$; if we ensure (5) by giving each gate a different $x$-coordinate, particles may have to travel a horizontal distance of $\mathcal{O}(n)$ to meet,

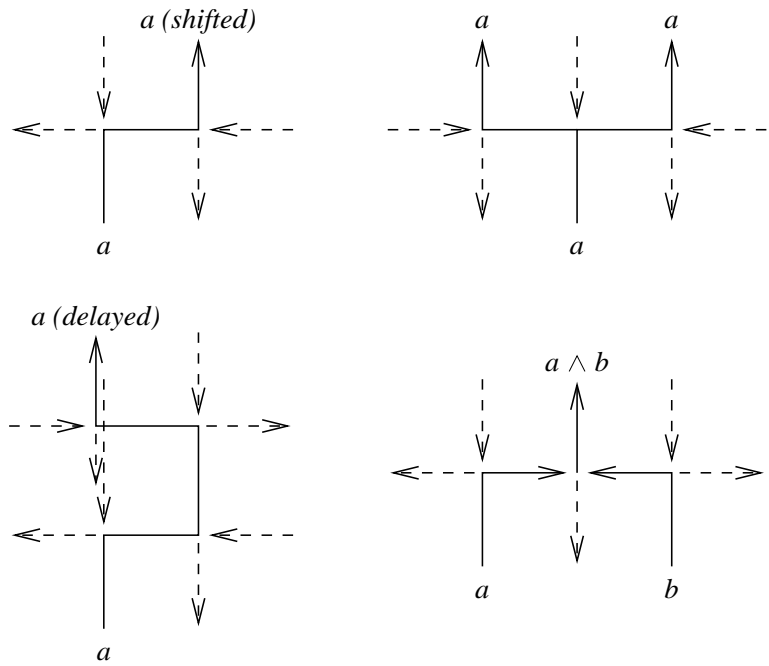

**Fig. 2.** Shifting, splitting, and delaying wires, and an AND gate, in the HPP lattice gas. Dashed arrows are part of the gate, or "exhaust" particles that may be produced; solid arrows are particles that carry truth values.

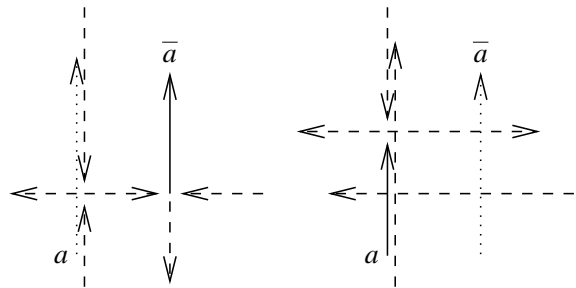

**Fig. 3.** Negation in the HPP lattice gas. If $a$ is not present then the two vertical dashed particles collide, producing a particle which collides with another approaching from the right, and $\overline{a}$ is produced; if $a$ is present (solid) it collides with the upper dashed particle, their output misses the left-moving particle, and $\overline{a}$ is not produced. A dotted arrow indicates a particle that isn't there.

so the total time is at most $\mathcal{O}(n^2)$.

In fact, we can improve this time bound by stratifying the circuit into levels and doing each level in parallel. Let the inputs be level 0, and let level $k+1$ consist of those gates whose inputs are at level $k$ or less. Then the *depth D* of the circuit is the level of the outputs, and its *width W* is the maximum number of gates at any one level.

We can do the AND gates and negations at a given level almost all at once, separating them by one site in the $y$-direction as shown in figure 4 so that sideways-moving particles don't collide with each other. Particles travel a horizontal distance $\mathcal{O}(W)$ to meet in their AND gates, after being synchronized by delays of $\mathcal{O}(W)$, and then move horizontally by $\mathcal{O}(W)$ so that the next level of gates will not interfere with this one. Thus the time and spatial extent of the CA we need is $\mathcal{O}(W)$ for each level; summing over all the levels gives a time and length $\mathcal{O}(DW) = \mathcal{O}(n)$.

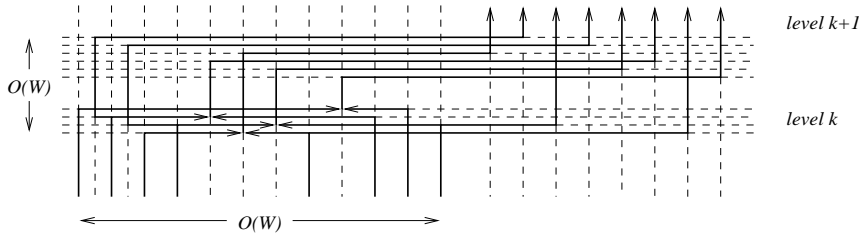

**Fig. 4.** Doing each level of the circuit at almost the same time, and shifting the wires to avoid interference with the next level. Delays are not shown.

Since the time and length for the lattice gas are linear in the number of gates, a Boolean circuit of polynomial size can be reduced to predicting a lattice gas for a polynomial amount of time and space. Thus we have shown that HPP LATTICE GAS PREDICTION is **P**-complete.[3]

## 3  The FHP III lattice gas

The FHP lattice gas lives on the triangular lattice. Unlike the HPP rule, it becomes isotropic in the continuum limit, making it an attractive way to simulate the Navier-Stokes equations. We use a variant called FHP III [4], which includes both stationary and moving particles; this allows particles to have a varying speed or energy, so that the system has non-trivial thermodynamic behavior. Collisions are shown in figure 5, except for head-on collisions, which non-deterministically rotate the outgoing particles by ±60°. We will not use head-on collisions in our simulation.

---

[3] Strictly speaking, for LATTICE GAS PREDICTION to be in **P** we need to express $t$ in unary, so that even for initial conditions with very few particles the size of the input is $t$ rather than $\mathcal{O}(\log t)$.

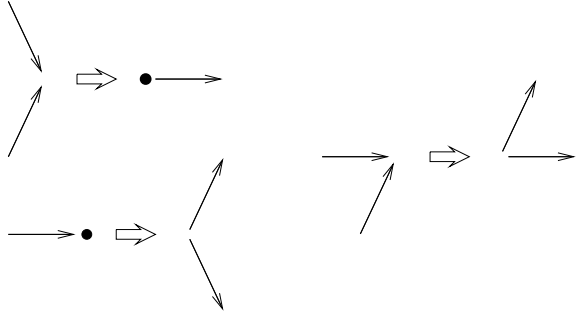

**Fig. 5.** Two-particle collisions in the FHP III lattice gas. Particles colliding at 120° produce a stationary particle and a moving one, and vice versa (conserving momentum but not energy), and particles colliding at 60° don't interact.

Then we can meet the same 5 criteria as before:

(1), (2) and (3) are shown in figure 6, where we use stationary particles to bend or split wires, and a 120° collision for AND.

(4) is shown in figure 7. Again, if $a$ is not present then a dashed particle sets off two collisions and produces $\overline{a}$, while if $a$ is present it collides with the first particle and prevents this from taking place.

(5) is slightly harder here than it was for the HPP rule, since diagonal and vertical exhaust can interact. Again, we stratify the circuit into $D$ levels, with at most $W$ gates in each. Then note the following facts:

a.) Upward-diagonal particles only interact with each other if they are at the same $y$ coordinate, i.e. if the displacement between them is horizontal, and similarly for downward-diagonal particles.

b.) Upward-diagonal and downward particles only interact if the displacement between them is 60° from the horizontal, and similarly for downward-diagonal and upward particles.

c.) Upward-diagonal and downward-diagonal particles only interact (through head-on collisions) if their displacement is 30° from the horizontal.

As shown in figure 8, all of these things can be avoided if we arrange gates in each level so that they are displaced from each other by less than 30°, but more than 0°, from the horizontal. This is easily accomplished within a rectangle whose height and width are linear in $W$.

Thus we can simulate circuits level-by-level as we did for the HPP lattice gas, using $\mathcal{O}(n)$ time and length for circuits with $n$ gates. So FHP III LATTICE GAS PREDICTION is **P**-complete also.

We note that a similar result for the FHP III rule was obtained in [16] under the name of "computation universality," but with additional barrier particles of infinite mass that other particles bounce off of elastically.

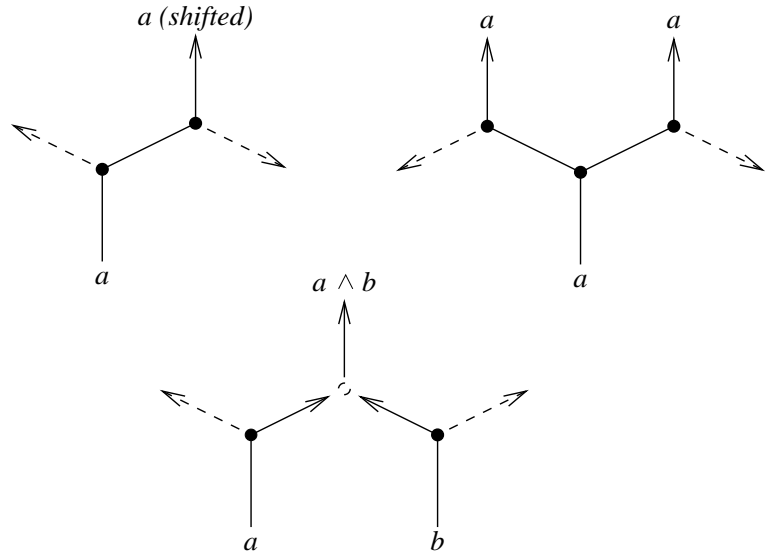

**Fig. 6.** Shifting and splitting wires, and an AND gate, in the FHP III lattice gas. Solid dots indicate stationary particles that are initially present; dashed ones may be produced as exhaust.

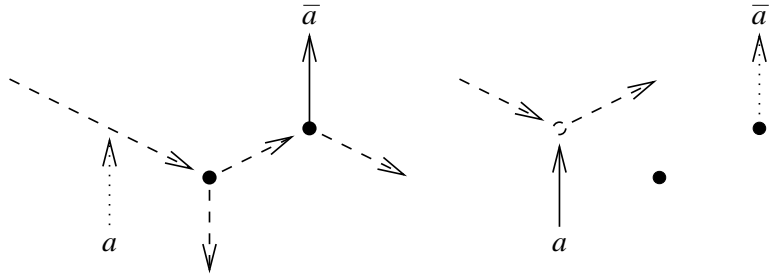

**Fig. 7.** Negation in the FHP III lattice gas.

## 4 Strict P-completeness and polynomial speedup

Since LATTICE GAS PREDICTION is **P**-complete, lattice gases cannot be predicted in polylogarithmic parallel time, unless **NC** = **P** and all problems in **P** can be parallelized to this extent. But even if $\mathbf{NC} \subsetneq \mathbf{P}$, it might still be possible to achieve a *polynomial* speedup with parallelization, predicting $t$ time-steps in parallel time $\mathcal{O}(t^\alpha)$ for some $\alpha < 1$. Such problems are in the class **SP** of *semi-efficient* parallel computation, introduced in [17].

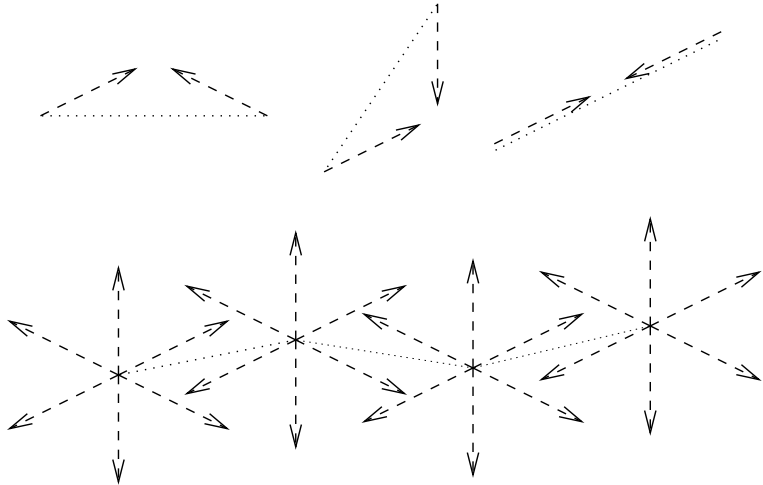

**Fig. 8.** Collisions between exhaust particles only take place if they are displaced by $0°$, $30°$ or $60°$ from the horizontal. We can avoid this by arranging gates in each level so that they have a small (less than $30°$ but more than $0°$) displacement in the $y$-direction.

Clearly even a polynomial speedup is highly useful from a computational point of view; $t^{1/2}$ is an enormous improvement over $t$ for large $t$. For instance, Moriarty, Machta and Greenlaw give an algorithm with $\alpha < 1$ on average for diffusion-limited aggregation [14]. They call $\alpha$ the *dynamic exponent*, and propose it as a measure of the extent to which a system's dynamics is dependent on its history.[4]

Is $\mathbf{SP} = \mathbf{P}$? That is, can all problems in $\mathbf{P}$ be parallelized with $\alpha < 1$? To study this, Condon introduced the notion of strict $\mathbf{P}$-completeness [1]. Although the precise definition is somewhat technical, roughly speaking a problem $\Pi$ is *at most $T(n)$-$\mathbf{P}$-complete* if, for any other problem $\Pi' \in \mathbf{P}$ that can be solved in serial time $t'(n)$, there is an $\mathbf{NC}$ reduction $f$ from $\Pi'$ to $\Pi$ such that $T(|f(n)|) = \mathcal{O}(t'(n))$ (plus a possible factor polylogarithmic in $n$). Then if we can solve $\Pi$ polynomially faster than $T(n)$ in parallel, then we can solve $\Pi'$ polynomially faster than $t'(n)$, and every problem in $\mathbf{P}$ has a polynomial speedup. If $\Pi$ in fact can be solved in parallel time $\mathcal{O}(T(n))$, then it is *strict $T(n)$-$\mathbf{P}$-complete*.

Condon's main result is that CIRCUIT VALUE is strict $n^{1/2}$-$\mathbf{P}$-complete for *square* circuits, whose width is equal to their depth; that is, that unless $\mathbf{SP} = \mathbf{P}$, the number of parallel steps needed is proportional to the circuit's depth $D = W = n^{1/2}$.

Our reduction from circuits to lattice gases can be done in $\mathbf{NC}$ since it consists of a local replacement of gates with groups of collisions. Furthermore, the time, size, and number of particles in the CA configuration it gives are all linear in $n$, so the input size is $|f(n)| = \mathcal{O}(n \log n)$ if we list each particle's initial position in a square of size $\mathcal{O}(n)$. Thus the reduction from square circuits to lattice gas states is linear (plus a logarithmic factor), and LATTICE GAS PREDICTION is at most $t^{1/2}$-**P**-complete. We can say then that lattice gases have a dynamic exponent $\alpha$ of at least $1/2$ unless $\mathbf{SP} = \mathbf{P}$.

---

[4] Moriarty and Machta define $\alpha$ in terms of the system's size; using time seems more appropriate for a CA.

In fact, we conjecture that lattice gases are strict $t$-**P**-complete and have a dynamic exponent of $\alpha = 1$, i.e. that they cannot be parallelized to better than linear time, so that explicit simulation is the best we can do. This is because the degree to which we can parallelize the system seems bounded by the mean free path of the particles, which is a function of density but stays roughly constant in time.

One way to prove this would be to show that, for each $\alpha < 1$, the family of circuits with $D = n^\alpha$ is strict $n^\alpha$-**P**-complete; then CIRCUIT VALUE would be strict $n$-**P**-complete in general. Unfortunately, the proof in [1] doesn't appear to work for circuits narrower than $\alpha = 1/2$.

## 5 Is all this physically relevant?

We have shown that LATTICE GAS PREDICTION for the HPP and FHP III rules is **P**-complete, and in fact at most $t^{1/2}$-**P**-complete. Thus it cannot be solved in parallel in polylogarithmic time, or even $t^\alpha$ for $\alpha < 1/2$, unless all problems in **P** are easier to solve than we think; in other words, explicit simulation is probably the only way to predict a lattice gas.

However, this begs the following question: what relevance, if any, do these results have to the complexity of the Navier-Stokes equation or continuum hydrodynamics? In the continuum limit we are interested in averages, not the presence or absence of a single particle at a single site. Our intuition about turbulence is that, as a highly non-linear system, it must be integrated numerically, and no short-cuts are possible. But we have not proved that here, as the following discussion will show.

If a fast algorithm existed for the lattice gas CA, then we could use it to predict the Navier-Stokes equation quickly: just convert the continuum state into a CA state, predict the CA, and average to get the continuum state back. Thus the complexity of a discrete system is an upper bound on the complexity of its continuum limit (assuming that choosing a CA state and taking averages can be done quickly).

However, the converse is not the case. The heat equation, for instance, is linear and easily predictable, even though it is the continuum limit of the computationally universal "billiard-ball" CA of Fredkin and Toffoli [2]. Thus the continuum limit of a CA can be much easier to predict than the CA itself, and a hardness result for the CA is not necessarily a hardness result for the corresponding continuum system.

It would be extremely valuable if a continuous analog of **P**-completeness, showing that numerical integration is the only way to predict the system, could be proved for a naturally occuring partial differential equation. We expect that this will require radically different techniques than for the discrete case. In the meantime, we hope that similar complexity results — either **P**-completeness proofs or efficient prediction algorithms — can be found for other cellular automata of physical interest.

**Acknowledgements.** We thank Jonathan Machta and Ken Moriarty for a careful reading of the manuscript, the Bellairs Research Institute in Barbados for their hospitality, and Rocky for cheese cutters and rum punch.

# References

1. A. Condon, "A theory of strict **P**-completeness." STACS 1992, in *Lecture Notes in Computer Science* **577** (1992) 33–44.
2. E. Fredkin and T. Toffoli, "Conservative logic." *Int. J. Theor. Phys.* **21** (1982) 219–253.
3. U. Frisch, B. Hasslacher, and Y. Pomeau, "Lattice-gas automata for the Navier-Stokes equation." *Phys. Rev. Lett.* **56** (1986) 1505–1508.
4. U. Frisch, D. d'Humieres, B. Hasslacher, P. Lallemand, Y. Pomeau, and J.P. Rivet, "Lattice gas hydrodynamics in two and three dimensions." *Complex Systems* **1** (1987) 649–707.
5. R. Greenlaw, H.J. Hoover, and W.L. Ruzzo, *Limits to Parallel Computation:* **P**-*Completeness Theory.* Oxford University Press, 1995.
6. J. Hardy, O. de Pazzis, and Y. Pomeau, "Molecular dynamics of a classical lattice gas: transport properties and time correlation functions." *Phys. Rev.* **A13** (1976) 1949–1960.
7. K. Lindgren and M.G. Nordahl, "Universal Computation in Simple One-Dimensional Cellular Automata." *Complex Systems* **4** (1990) 299–318.
8. J. Machta and R. Greenlaw, "The computational complexity of generating random fractals." *J. Stat. Phys.* **82** (1996) 1299.
9. C. Moore, "Quasi-linear cellular automata." Santa Fe Institute Working Paper 95-09-078, to appear in *Physica D*, *Proceedings of the International Workshop on Lattice Dynamics.*
10. C. Moore, "Non-Abelian cellular automata," Santa Fe Institute Working Paper 95-09-081, or "Predicting non-linear cellular automata quickly by decomposing them into linear ones", available as patt-sol/9701008. Submitted to *Physica D.*
11. C. Moore, "Majority-Vote Cellular Automata, Ising Dynamics, and **P**-Completeness." Santa Fe Institute Working Paper 96-08-060, to appear in *J. Stat. Phys.*
12. C. Moore, "The computational complexity of sandpiles." In preparation.
13. J. Machta and K. Moriarty, "The computational complexity of the lorentz lattice gas." Available as comp-gas/9607001, to appear in *J. Stat. Phys.*
14. K. Moriarty, J. Machta and R. Greenlaw, "Parallel algorithm and dynamic exponent for diffusion-limited aggregation." Available as comp-gas/9609001, to appear in *Phys. Rev. E.*
15. C.H. Papadimitriou, *Computational Complexity.* Addison-Wesley, 1994.
16. R. Squier and K. Steiglitz, "Two-dimensional FHP lattice gases are computation universal." *Complex Systems* **7** (1993) 297–307.
17. J.S. Vitter and R.A. Simons, "New classes for parallel complexity: a study of unification and other complete problems for **P**." *IEEE Transactions on Computers* **TC-35** (1986) 403–418.

This article was processed using the LaTeX macro package with LLNCS style